\documentstyle[epsfig]{aipproc}

\begin{document}

%\hspace*{10cm} June 2000

\hspace*{10cm} ADP-00-32/T415

\hspace*{10cm} JLAB-THY-00-22	\\

\title{Asymmetric Quarks in the Proton%}
\thanks{Talk presented at the 3rd International Symposium on
Symmetries in Subatomic Physics, Adelaide, Australia, March 2000.}}

\author{W. Melnitchouk}

\address{Special Research Centre for the
        Subatomic Structure of Matter,  \\ 
        University of Adelaide, Adelaide 5005, Australia, and\\
	Jefferson Lab, 12000 Jefferson Avenue,
        Newport News, VA 23606}

\maketitle

\begin{abstract}
Asymmetries in the quark momentum distributions in the proton reveal
fundamental aspects of strong interaction physics.
Differences between $\bar u$ and $\bar d$ quarks in the proton sea
provide insight into the dynamics of the pion cloud around the nucleon
and the nature of chiral symmetry breaking.
Polarized flavor asymmetries allow the effects of pion clouds to be
disentangled from those of antisymmetrization.
Asymmetries between $s$ and $\bar s$ quark distributions in the nucleon
are also predicted from the chiral properties of QCD.
\end{abstract}

%%%%%%%%%%%%%%%%%%%%%%%%%%%%%%%%%%%%%%%%%%%%%%%%%%%%%%%%%%%%%%%%%%%%%%%%%%
\section*{Introduction}

Asymmetries in the proton's spin and flavor quark distributions
provide direct information about QCD dynamics of bound systems.
Differences between quark and antiquark distributions in the proton sea
almost universally signal the presence of phenomena which require
understanding of strongly coupled QCD.
Their existence testifies to the relevance of long-distance dynamics
% (which are responsible for confinement)
even at large energy and momentum transfers.

Over the past decade a number of high-energy experiments and refined
data analyses have forced a re-evaluation of our view of the nucleon
in terms of three valence quarks immersed in a sea of perturbatively   
generated $q\bar q$ pairs and gluons~\cite{SCIENCE}.
A classic example of this is the asymmetry of the light quark sea of  
the proton, dramatically confirmed in recent deep-inelastic and Drell-Yan
experiments at CERN~\cite{NMCGSR,NA51} and Fermilab~\cite{E866}.
Less firmly established, but equally intriguing, are asymmetries
between quark and antiquark distributions for heavier flavors,
such as $s$ and $\bar s$, which can be measured in deep-inelastic
neutrino scattering experiments~\cite{CCFR}, or even $c$ and
$\bar c$~\cite{MTC,SMT,PAIVA,BROD}.

%%%%%%%%%%%%%%%%%%%%%%%%%%%%%%%%%%%%%%%%%%%%%%%%%%%%%%%%%%%%%%%%%%%%%%%%%%
\section*{Light Antiquark Asymmetry}

Because gluons in QCD are flavor-blind, the sea generated through the
perturbative process $g \rightarrow q \bar q$ is symmetric in the quark
flavors.
Differences can arise due to different quark masses, but because isospin
symmetry is such a good symmetry in nature, one expects that the sea of
light quarks generated perturbatively would be almost identical,
$\bar u(x) = \bar d(x)$.

It was therefore a surprise to many when measurements by the New Muon
Collaboration (NMC) at CERN~\cite{NMCGSR} of the proton -- neutron
structure function difference suggested a significant excess of $\bar d$
over $\bar u$ in the proton.
Indeed, it heralded a renewed interest in the application of ideas 
from non-perturbative QCD to deep-inelastic scattering analyses.
While the NMC experiment measured the integral of the antiquark
difference, more recently the E866 Collaboration at Fermilab has for
the first time mapped out the shape of the $\bar d / \bar u$ ratio
over a large range of $x$, $0.02 < x < 0.345$.

Specifically, the E866/NuSea Collaboration measured $\mu^+\mu^-$
Drell-Yan pairs produced in $pp$ and $pd$ collisions.
If $x_1$ and $x_2$ are the light-cone momentum fractions carried by
partons in the projectile and target, respectively, then in the limit
$x_1 \gg x_2$ the ratio of $pd$ and $pp$ cross sections can be
written~\cite{E866}:
\begin{eqnarray}
\label{sigdy}
{ \sigma^{pd} \over 2 \sigma^{pp} }
&=& {1 \over 2}
\left( 1 + { \bar d(x_2) \over \bar u(x_2) } \right)
{ 4 + d(x_1)/u(x_1) \over
  4 + d(x_1)/u(x_1) \cdot \bar d(x_2)/\bar u(x_2) }\ ,
\end{eqnarray}
where isospin symmetry has been used to relate quark distributions
in the neutron to those in the proton.
Corrections for nuclear shadowing in the deuteron~\cite{MTSHAD},
which are important at small $x$, are negligible in the region covered
by this experiment.

The relatively large asymmetry found in these experiments, shown in
Fig.~1, implies the presence of non-trivial dynamics in the proton
sea which does not have a perturbative origin.
The simplest and most obvious source of a non-perturbative asymmetry
in the light quark sea is the chiral structure of QCD.

{}From numerous studies in low energy physics, including chiral
perturbation theory, pions are known to play a crucial role in the
structure and dynamics of the nucleon~\cite{CBM}.
%
% and there is no reason
% why the long-range tail of the nucleon should not also play a role at
% higher energies.
%
As pointed out by Thomas~\cite{AWT83}, if the proton's wave function
contains an explicit $\pi^+ n$ Fock state component, a deep-inelastic
probe scattering from the virtual $\pi^+$, which contains a valence
$\bar d$ quark, will automatically lead to a $\bar d$ excess in the 
proton.
In the impulse approximation, deep-inelastic scattering from the $\pi N$
component of the proton can then be understood in the infinite momentum
frame (IMF)~\cite{IMF} as the probability for a pion to be emitted by the
proton, folded with the probability of finding the a parton in the pion.
For the antiquark asymmetry, this can be written
as~\cite{SULL,ZOL,MTV,DYN,REV}:
\begin{eqnarray}
\label{conv}
\bar d(x) - \bar u(x)
&=& {2 \over 3} \int_x^1 {dy \over y}
    f_{\pi N}(y)\ \bar d^{\pi^+} (x/y)\ ,
\end{eqnarray}
where $\bar d^{\pi^+}$ is the (valence) $\bar d$ quark distribution
in the $\pi^+$, and the distribution of pions with a recoiling nucleon
($N \rightarrow \pi N$ splitting function) is given
by~\cite{AWT83,SULL,ZOL,MTV,DYN,REV}:
\begin{eqnarray}
\label{fypin}
f_{\pi N}(y)
&=&
{ 3 g^2_{\pi N N} \over 16 \pi^3 }\
\int{ d^2{\bf k}_T \over (1-y) }
\frac{ {\cal F}_{\pi N}^2(s_{\pi N}) }
     { y\ (M^2 - s_{\pi N})^2 }
\left( { k_T^2 + y^2 M^2 \over 1-y } \right)\ ,
\end{eqnarray}
where $s_{\pi N}$ is the invariant mass squared of the $\pi N$ system,
$s_{\pi N} = (k^2_T + m_{\pi}^2)/y + (k^2_T + M^2)/(1-y)$, and the
$\pi NN$ vertex form factor, ${\cal F}_{\pi N}$, plays the role of an
ultraviolet cut-off.

Another contribution known to be important for nucleon structure
is that from the $\pi \Delta$ component of the nucleon wave
function~\cite{CBM}.
For a proton initial state, the dominant Goldstone boson fluctuation is
$p \to \pi^- \Delta^{++}$, which leads to an excess of $\bar u$ over
$\bar d$.
The relative contributions of the $\pi N$ and $\pi\Delta$ components
are determined partly by the $\pi NN$ and $\pi N\Delta$ vertex form
factors.
The most direct way to constrain these is through a comparison with
the axial form factors for the nucleon and for the $N$--$\Delta$
transition~\cite{AXIAL}.
Within the framework of PCAC the axial form factors are directly
related to those corresponding to pion emission or absorption.
The data on the axial form factors are best fit, in a dipole
parameterization, by a 1.3~(1.02)~GeV dipole for the axial $N$
($N$--$\Delta$ transition) form factor~\cite{AXIAL}, which gives
a pion probability in the proton of $\approx$ 13\% (10\%).

\begin{figure}[ht]
\centerline{\epsfig{file=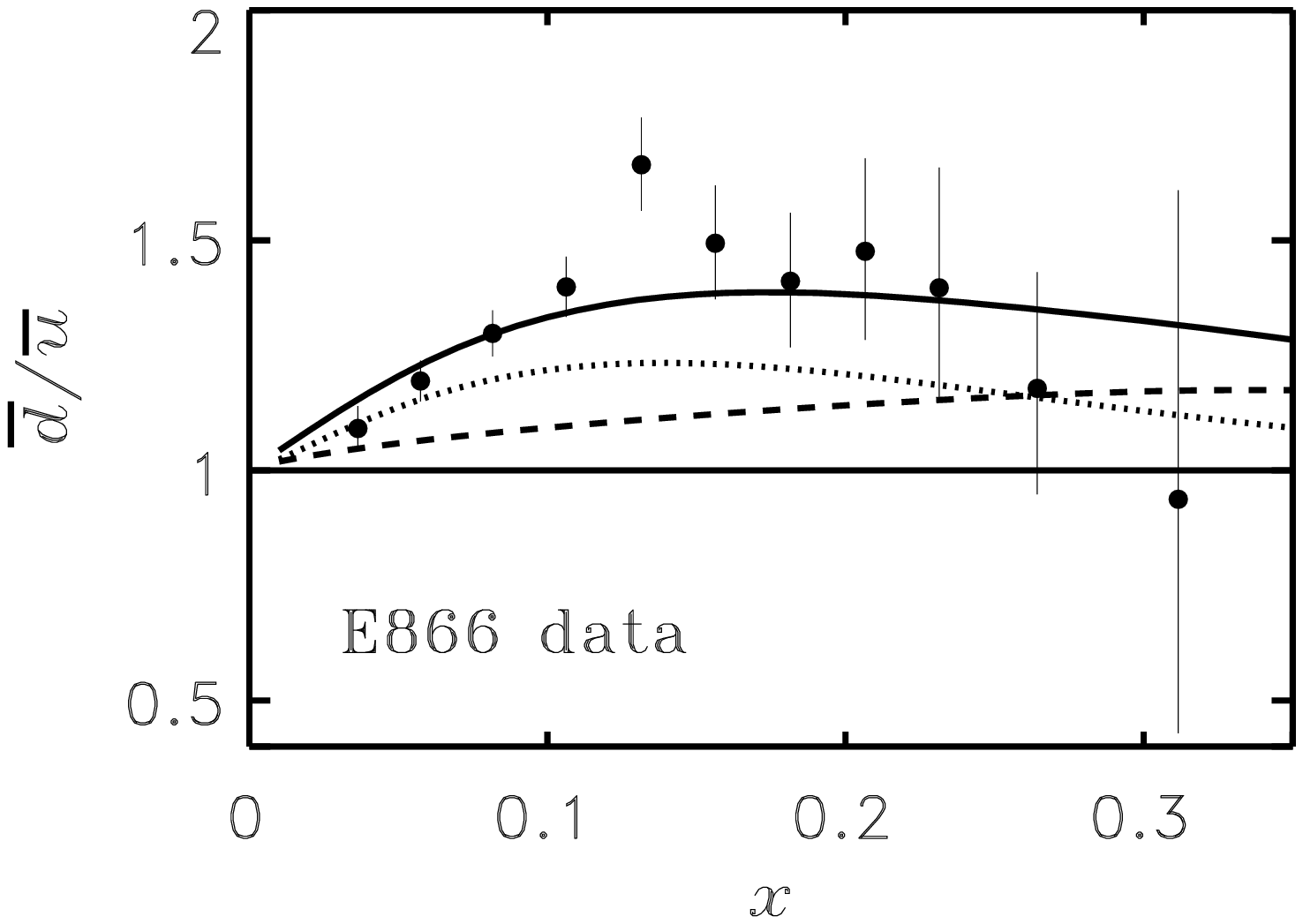,height=6cm}}
% \vspace{10pt}
\caption{Flavor asymmetry of the light antiquark sea, including pion
        cloud (dashed) and Pauli blocking effects (dotted), and the
        total (solid)~\protect\cite{DYN}.}
\label{fig1}
\end{figure}

The resulting $\bar d/\bar u$ ratio
% calculated from the pion cloud~\cite{DYN},
is shown in Fig.~1 (dashed line).
Data on the sum of $\bar u$ and $\bar d$ (which is dominated by
perturbative contributions) have been used to convert the calculated
$\bar d-\bar u$ difference to the $\bar d/\bar u$ ratio.
The results suggest that with pions alone one can account for about half
of the observed asymmetry, leaving room for possible contributions from
other mechanisms.
One can fine tune the cut-off parameters, or include other, heavier
mesons and baryons in the proton's cloud~\cite{REV} to obtain a better
fit, however, the fact that an asymmetry exists follows directly from
the chiral properties of QCD.

In particular, one can derive the non-analytic behavior of flavor
asymmetries in the nucleon sea by considering the chiral
($m_\pi \rightarrow 0$) limit of Goldstone boson loops.
The leading non-analytic (LNA) behavior of the excess number of
$\bar{d}$ over $\bar{u}$ quarks in the proton has a chiral behavior
typical of loop expansions in chiral effective theories, such as chiral
perturbation theory~\cite{LNA}:
\begin{eqnarray}
\label{result}
\int_0^1 dx \left( \bar{d}(x) - \bar{u}(x) \right)_{\rm LNA}
&=& { 2 g_A^2 \over (4 \pi f_\pi)^2 }\
m_\pi^2  \log (m_\pi^2/\mu^2)\ ,
\end{eqnarray}
where $\mu$ is an ultraviolet cut-off mass,
and the PCAC relation has been used to express the $\pi NN$ coupling
constant in terms of the axial charge, $g_A$.
This result also generalizes to higher moments, each of which has a
non-analytic component, so that the $\bar d - \bar u$ distribution
itself, as a function of $x$, has a model-independent, LNA component.
The presence of non-analytic terms indicates that Goldstone bosons  
play a role which cannot be canceled by any other physical process
(except by chance at a particular value of $m_\pi$).

Another mechanism which could also contribute to the $\bar d-\bar u$
asymmetry is associated with the effects of antisymmetrization of
$q \bar q$ pairs created inside the nucleon~\cite{SST,SMST,ST}.
As pointed out originally by Field and Feynman~\cite{FF}, because the
valence quark flavors are unequally represented in the proton, the Pauli
exclusion principle will affect the likelihood with which $q\bar q$
pairs can be created in different flavor channels.
Since the proton contains two valence $u$ quarks compared with only one
valence $d$ quark, $u \bar u$ pair creation will be suppressed relative
to $d \bar d$ creation.
In the ground state of the proton the suppression will be in the
ratio $\bar d : \bar u = 5:4$~\cite{SST}.

The shape of the Pauli contribution to the asymmetry is difficult to
predict model-independently, but is expected to have an $x$ dependence
typical of sea quark distributions~\cite{SST}.
Phenomenologically, one finds a good fit for $x < 0.2$ if roughly half
of the asymmetry is attributed to antisymmetrization~\cite{DYN,MTS}.
At larger $x$ it is difficult to reproduce the apparent trend in the
data towards zero asymmetry, and possibly even an excess of $\bar u$
for $x > 0.3$.
Unfortunately, the error bars are quite large beyond $x \sim 0.25$,
and it is not clear whether new Drell-Yan data will be forthcoming
in the near future to clarify this.

A solution may be available, however, through semi-inclusive
scattering, in which one tags charged pions produced off protons and
neutrons~\cite{LMS}.
The HERMES Collaboration has in fact recently measured this
ratio~\cite{HERMES}, although there the rapidly falling cross sections
at large $x$ make measurements beyond $x \sim 0.3$ challenging.
On the other hand, a high luminosity electron beam such as that    
available at Jefferson Lab could allow a precise measurement of the
asymmetry beyond $x \sim 0.3$~\cite{JIANG}.

The relative roles of pions and Pauli blocking can be further
disentangled by measuring the polarized flavor distributions
$\Delta\bar d$ and $\Delta\bar u$ in semi-inclusive
scattering~\cite{POLSEA}.
While the antiquark sea due to pions is necessarily unpolarized,
the Pauli exclusion principle predicts quite a large asymmetry,
$\Delta \bar u : \Delta \bar d = -4:1$
in quark models with SU(6) symmetry~\cite{SST,SMST}.
By fixing the normalization of the Pauli effect from the polarized
asymmetries, one could then determine the magnitude of the pion cloud
contribution to $\bar d-\bar u$.

%%%%%%%%%%%%%%%%%%%%%%%%%%%%%%%%%%%%%%%%%%%%%%%%%%%%%%%%%%%%%%%%%%%%%%%%%%
\section*{Strangeness in the Nucleon}

A complication in studying the light quark sea is the fact that
non-perturbative features associated with $u$ and $d$ quarks are
intrinsically correlated with the valence core of the proton,
so that effects of $q \bar q$ pairs can be difficult to distinguish
from those of antisymmetrization, or residual interactions of qurks
in the core.
The strange sector, on the other hand, where antisymmetrization between
sea and valence quarks plays no role, is therefore likely to provide
even more direct information about the non-perturbative origin of the
nucleon sea~\cite{JT}.

Generalizing the formal LNA analysis to the flavor SU(3) sector, one
can show that the existence of an asymmetry between $s$ and $\bar s$
quarks in the nucleon is predicted on the basis of chiral SU(3) symmetry
breaking, which gives rise to a kaon cloud through the fluctuation
$N \rightarrow K Y$, where the hyperon $Y = \Lambda, \Sigma, \cdots$.
Since the $\bar s$ quark typically comes from the $K$ and the $s$ from
the hyperon, the different $K$ and $Y$ masses and momentum distributions
naturally lead to a difference between the $s$ and $\bar s$ distributions
in the nucleon~\cite{ST_STRANGE,GI}.
For the case of the $\Lambda$, one has~\cite{ST_STRANGE,MM}:
\begin{eqnarray}
s(x) - \bar s(x)
&=& \int_x^1 { dy \over y }
\left( f_{\Lambda K}(y)\ s^{\Lambda}(x/y)\
     - f_{K \Lambda}(y)\ \bar s^K(x/y)
\right)\ ,
\end{eqnarray}
where $f_{K\Lambda}$ is the analog of the $\pi N$ splitting function
in Eq.(\ref{fypin}), and $f_{\Lambda K}(y) = f_{K\Lambda}(1-y)$.
Zero net strangeness in the nucleon implies the vanishing of the lowest
moment of $s-\bar s$, however, higher moments in general do not vanish.
In particular, the LNA behavior of the $n$-th moment of the
$N \to K \Lambda$ splitting function is~\cite{LNA}:
\begin{eqnarray}
\left. f_{K\Lambda}^{(n)} \right|_{\rm LNA}
% &=& \left( { g_{KN\Lambda}^2 \over 16 \pi^2 } \right)
&=& { 27 \over 25 } { M^2 g_A^2 \over (4 \pi f_\pi)^2 }
(M_\Lambda-M)^2 (-1)^n { m_K^{2n+2} \over \Delta M^{2n+4} }
\log (m_K^2/\mu^2)\ ,
\end{eqnarray}
where $\Delta M^2 = M_\Lambda^2-M^2$,
and SU(6) symmetry has been used to relate $g_{KN\Lambda}$ to $g_A/f_\pi$.
Since the LNA terms in the chiral expansion are in general not canceled
by other contributions, the process of dynamical symmetry breaking in
QCD implies that the $s$ and $\bar s$ distributions must therefore have
a different dependence on Bjorken $x$.

\begin{figure}[ht]
\centerline{\epsfig{file=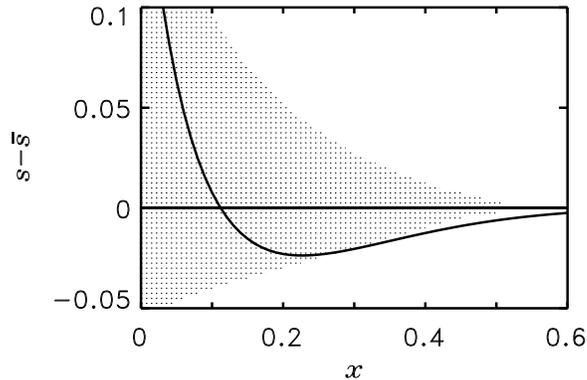,height=6cm}}
\caption{Strange quark asymmetry in the proton arising from a
	kaon cloud of the nucleon~\protect\cite{MM}, with $\approx$ 3\%
	kaon probability.
	The shaded region indicates current experimental limits
	from the CCFR Collaboration~\protect\cite{CCFR}.}
\label{fig2}
\end{figure}

The $s$ and $\bar s$ distributions can be individually measured in
charm production in deep-inelastic $\nu$ and $\bar \nu$ scattering.
Unfortunately, because of relatively large errors the available data
from the CCFR collaboration~\cite{CCFR}, indicated by the shaded region
in Fig.~2, are unable to distinguish between zero asymmetry and a small
amount of non-perturbative strangeness as would be expected from kaon
loops~\cite{MM} (solid line in Fig.~2).
More precise data would be valuable in determining the magnitude,
and even the sign, of the asymmetry as a function of $x$, which depends
rather strongly on the dynamics of the $KNY$ interaction~\cite{MM}.

\newpage
%%%%%%%%%%%%%%%%%%%%%%%%%%%%%%%%%%%%%%%%%%%%%%%%%%%%%%%%%%%%%%%%%%%%%%%%%
\section*{Conclusion}

We have outlined a number of specific examples where measurement of
asymmetries in the sea quark distributions of the nucleon can reveal
hitherto hidden details of its non-perturbative structure.
Asymmetries are predicted to exist on general grounds from the chiral
properties of QCD, by examining the leading non-analytic chiral behavior
of quark distributions associated with Goldstone boson loops.

For the $\bar d/\bar u$ ratio, it is important experimentally to confirm
the downward trend of the ratio at large $x$, which may be feasible
through semi-inclusive $\pi^\pm$ production.
Interestingly, Goldstone boson loops do not give rise to any flavor
asymmetries for spin-dependent antiquark distributions, which can only
arise from Pauli blocking effects in the proton.

For the strange content of the nucleon, the data from CCFR continue
to be reanalyzed in view of possible nuclear shadowing corrections
and charm quark effects~\cite{BOROS} on $s$ and $\bar s$.
Together with complementary data on strange form factors currently
being gathered in parity-violating elastic electron scattering 
experiments~\cite{PARITY}, this will provide valuable information
about the role of non-perturbative strangeness in the nucleon.	\\

This work was supported by the Australian Research Council and
DOE contract \mbox{DE-AC05-84ER40150}.

\newpage
%%%%%%%%%%%%%%%%%%%%%%%%%%%%%%%%%%%%%%%%%%%%%%%%%%%%%%%%%%%%%%%%%%%%%%%%%%%


\begin{references}

\bibitem{SCIENCE}
Watson, A.,
{\em Science} {\bf 283}, 472 (1999).

\bibitem{NMCGSR}
Amaudraz, P., et al.,
{\em Phys. Rev. Lett.} {\bf 66}, 2712 (1991).

\bibitem{NA51}
Baldit, A., et al.,
{\em Phys. Lett.} {\bf B 332}, 244 (1994).

\bibitem{E866}
Hawker, E.A., et al.,
{\em Phys. Rev. Lett.} {\bf 80}, 3715 (1998).

\bibitem{CCFR}
Bazarko, A.O., et al.,
{\em Zeit. Phys.} {\bf C 65}, 189 (1995).

\bibitem{MTC}
Melnitchouk, W. and Thomas, A.W.,
{\em Phys. Lett.} {\bf B 414}, 134 (1997).

\bibitem{SMT}
Steffens, F.M., Melnitchouk, W. and Thomas, A.W.,
{\em Eur. Phys. J.} {\bf C 11}, 673 (1999).

\bibitem{PAIVA}
Paiva, S., Nielsen, M., Navarra, F.S., Duraes, F.O. and Barz, L.L.,
{\em Mod. Phys. Lett.} {\bf A 13}, 2715 (1998).

\bibitem{BROD}
Brodsky, S.J. and Ma, B.Q.
{\em Phys. Lett.} {\bf B 381}, 317 (1996).

\bibitem{MTSHAD}
Melnitchouk, W. and Thomas, A.W.,
{\em Phys. Rev.} {\bf D 47}, 3783 (1993).

\bibitem{CBM}
Thomas, A.W.,
{\em Adv. Nucl. Phys.} {\bf 13}, 1 (1984).

\bibitem{AWT83}
Thomas, A.W.,
{\em Phys. Lett.} {\bf B 126}, 97 (1983).

\bibitem{IMF}
Drell, S.D., Levy, D.J. and Yan, T.M.,
{\em Phys. Rev.} {\bf D 1}, 1035 (1970).

\bibitem{SULL}
Sullivan, J.D.,
{\em Phys. Rev.} {\bf D 5}, 1732 (1972).

\bibitem{ZOL}
Zoller, V.R.,
{\em Z. Phys.} {\bf C 54}, 425 (1992).

\bibitem{MTV}
Melnitchouk, W. and Thomas, A.W.,
{\em Phys. Rev.} {\bf D 47}, 3794 (1993).

\bibitem{DYN}
Melnitchouk, W., Speth, J. and Thomas, A.W.,
{\em Phys. Rev.} {\bf D 59}, 014033 (1999).

\bibitem{REV}
Speth, J. and Thomas, A.W.,
{\em Adv. Nucl. Phys.} {\bf 24}, 83 (1998).

\bibitem{AXIAL}
Jones, G.T., et al.,
{\em Z. Phys.} {\bf C 43}, 527 (1989).

\bibitem{LNA}
Thomas, A.W., Melnitchouk, W. and Steffens, F.M.,
hep-ph/0005043.

\bibitem{SST}
Schreiber, A.W., Signal, A.I. and Thomas, A.W.,
{\em Phys. Rev.} {\bf D 44}, 2653 (1991).

\bibitem{SMST}
Schreiber, A.W., Mulders, P.J., Signal, A.I. and Thomas, A.W.,
{\em Phys. Rev.} {\bf D 45}, 3069 (1992).

\bibitem{ST}
Steffens, F.M. and Thomas, A.W.,
{\em Phys. Rev.} {\bf C 55}, 900 (1997).

\bibitem{FF}
Field, R.D. and Feynman, R.P.,
{\em Phys. Rev.} {\bf D 15}, 2590 (1977).

\bibitem{MTS}
Melnitchouk, W., Thomas, A.W. and Signal, A.I.,
{\em Z. Phys.} {\bf A 340}, 85 (1991).

\bibitem{LMS}
Levelt, J., Mulders, P.J. and Schreiber, A.W.,
{\em Phys. Lett.} {\bf B 263}, 498 (1991).

\bibitem{HERMES}
Ackerstaff, K., et al.,
{\em Phys. Rev. Lett.} {\bf 81}, 5519 (1998).

\bibitem{JIANG}
Jiang, X., et al.,
Jefferson Lab proposal P-00-115 (2000).

\bibitem{POLSEA}
Dressler, B., Goeke, K., Polyakov, M.V. and Weiss, C.,
{\em Eur. Phys. J.} {\bf C 14}, 147 (2000);
% 
Bhalerao, R.S.,
hep-ph/0003075.

\bibitem{JT}
Ji, X. and Tang, J.,
{\em Phys. Lett.} {\bf B 362}, 182 (1995).

\bibitem{ST_STRANGE}
Signal, A.I. and Thomas, A.W.,
{\em Phys. Lett.} {\bf B 191}, 206 (1987).

\bibitem{GI}
Geiger, P. and Isgur, N.,
{\em Phys. Rev.} {\bf D 55}, 299 (1997).

\bibitem{MM}
Melnitchouk, W. and Malheiro, M.,
{\em Phys. Rev. } {\bf C 55}, 431 (1997);
{\em Phys. Lett.} {\bf B 451}, 224 (1999).

\bibitem{BOROS}
Boros, C., Steffens, F.M., Londergan, J.T. and Thomas, A.W.,
{\em Phys. Lett.} {\bf B 468}, 161 (1999).

\bibitem{PARITY}
Mueller, B. et al.,
{\em Phys. Rev. Lett.} {\bf 78}, 3824 (1997);
%
Aniol, K.A. et al.,
{\em Phys. Rev. Lett.} {\bf 82}, 1096 (1999).

\end{references}
\end{document}